\documentstyle[11pt,newpasp,twoside,epsf]{article}
\begin{document}
\title{Sgr A* and Sgr A East: Intimate Life in the Galactic Center}
 \author{Tomasz Plewa}
\affil{Center for Astrophysical Thermonuclear Flashes,
The University of Chicago,
Chicago, IL 60637, USA}
\author{Micha\l\ R\'o\.zyczka}
\affil{N. Copernicus Astronomical Center,
Bartycka 18, 00716 Warsaw, Poland}

\begin{abstract}
A hydrodynamic interaction between an old supernova remnant and a
massive black hole is studied with the help of a high-resolution
adaptive mesh refinement code. A series of multidimensional numerical
models is obtained for a fixed distance between supernova explosion
site and the black hole but different densities of the ambient
medium. The interaction between the shell of the supernova remnant and
the black hole results in a rapid increase of the emission from the
vicinity of the black hole. Light curves of the gas heated by
gravitational compression and hydrodynamical interactions are obtained
assuming optically thin conditions. It is shown that the duration of
the event and the peak luminosity critically depend on the density of
the ambient medium.  The model is applied to the data obtained by the
Chandra satellite for the Galactic center.
\end{abstract}

\section{Introduction}

For several decades the structure and the contents of the Galactic
center (GC) region remained a mystery kept away from the eyes of
astronomers by dust extinction reaching $\approx$30 magnitudes in
optical part of the spectrum (Becklin et al. 1978). The earliest
studies in the infrared (Stebbins \& Whitford 1947) revealed an excess
in the emission in the direction of the GC but provided no information
about morphology.  Several authors (Hagen \& McClain 1954; McGee \&
Bolton 1954) speculated that a compact radio source located near the
Galactic Center, Sgr A, may actually be associated with the nucleus of
the Milky Way. Radio observations by McClain (1955) and Heeschen
(1955) provided first indication of the complex nature of Sgr A.

The discovery of a sub-parsec scale radio source in the GC region
(Balick \& Brown 1974) provided initial support for idea that galaxies
may harbor massive black holes in their nuclei (Lynden-Bell \& Rees
1971; Rees 1974).  High-resolution data obtained with the VLA (Brown,
Jonston, \& Lo 1981) and VLBI (Lo et al. 1981) allowed for
identification of the compact non-thermal radio source, Sgr A*, as the
primary candidate for the central massive object (Lo 1986). The VLA
also provided first detailed information about distribution of the gas
in the central ~10 pc revealing existence of a three-armed
``mini-spiral'' surrounding the central source (Ekers et al. 1983; Lo
\& Claussen 1983).

Observations obtained with the VLBA (Reid et al. 1999) and VLA (Backer
\& Sramek 1999) allowed to estimate the proper motion of the Sgr
A*. This data alone rules out stellar origin of the central object
and, if combined with the apparent size of Sgr A* (Lo et al. 1998;
Krichbaum, Witzel \& Zensus 1999; Doeleman et al. 2001), implies that
its mass has to be $> 10^4$ M\sun. More precise mass estimates based
on near-infrared observations of stellar motions in the central region
(Eckart \& Genzel 1996; Genzel et al. 1997; Ghez et al. 1998; Eckart
et al. 2002) yielded a value of $\approx 2.6\times10^6$ M\sun.

The X-ray emission from the GC region has been originally detected by
the Uhuru satellite (Kellog et al. 1971). A decade later Watson et
al. (1981) found an extended emission from the central region and
obtained the upper limit for the soft X-ray luminosity of the compact
object. The extended emission was further studied with Ginga (Yamauchi
et al. 1990), ASCA (Koyama et al. 1996), and Chandra (Bamba et
al. 2002). GC imaging with BeppoSAX (Sidoli et al. 1999), ASCA (Sakano
et al. 2002) and Chandra (Baganoff et al. 2002) allowed for resolving
a part of the extended emission in a number of discrete sources.

Finally, the GC region is known to be a source of hard X-ray (Skinner
et al. 1987; Churazov et al. 1994) and $\gamma$-ray emission
(von~Ballmoos, Diehl, \& Sch\"onfelder 1987; Mayer-Hasselwander et
al. 1998), but no evidence for the emission from Sgr A* itself has
been found so far. For more information about GC observations see
reviews by Genzel and Townes (1987), Yusef-Zadeh, Melia, \& Wardle
(2000), and Falcke and Melia (2001).

The most puzzling property of the GC black hole candidate, Sgr A*, is
its relative quiescence across the electromagnetic spectrum. However,
it might have looked different in the past. A discovery of fluorescent
X-ray emission from cold molecular clouds in the central region of the
Galaxy together with the apparent lack of irradiating source suggests
that not long ago Sgr A* might have been a bright source of
high-energy photons (Sunyaev, Markevitch, \& Pavlinsky 1993; Koyama et
al. 1996).  More recently, Murakami, Koyama, \& Maeda (2001) found
strong support for this idea and concluded that the Sgr B2 molecular
cloud could have been irradiated by a transient source located at the
very center of the Galaxy.

Maeda et al. (2002) argue that a likely cause of the last
high-luminosity event was an interaction between Sgr A* and the
supernova remnant Sgr A East.  In the present communication we report
preliminary results of hydrodynamical simulations related to this
possibility.

\section{Results and discussion}

Numerical models were obtained with the help of the AMRA code (Plewa
\& M\"uller 2001) which utilizes an adaptive mesh refinement method
and a high-order scheme to solve hydrodynamic equations.

The central source (CS), representing Sgr A* black hole, is a point
mass $m_{CS} = 2.6\times 10^{6}M\sun$, surrounded by an accreting
region of radius $r_{CS}$. In most cases we employ
$r_{CS}=4\times\Delta z$, where $\Delta z$ is the resolution of the
finest grid ($\Delta z\approx 2.8\times 10^{14}$ cm).

The supernova explosion energy, $E = 1\times 10^{51}$ erg, is constant
for all models. Since the parameters of the interstellar medium at the
GC are not well-known, runs with ambient density $n_H$ of $10^2$,
$3.16\times10^2$ and $10^3$ cm$^{-3}$ are performed. The ambient
pressure is the same in all models, and its value corresponds to a
temperature $T_{amb} = 10^4$ K in the $n_H=10^3$ model. Optically thin
radiative cooling with metallicity 4 times larger than the solar one
is employed.

For each density of the ambient medium we calculate a 1-D model of the
SNR with the resolution $\Delta r\approx 4.5\times 10^{15}$ cm. Its
evolution is followed for $10^4$ years, and then the model is mapped
onto a 2-dimensional cylindrical grid. At the same time, the CS is
positioned at the symmetry axis, in front of the remnant shell, and
approximately 0.5 pc away from it.

Table 1
%
%
\begin{table}
\begin{center}
\caption{Selected parameters of 1- and 2-D models at $t=10^4$ yr.}
\begin{tabular}{ccc}
&&\\
\tableline
$\log n_{amb}$ [$m_{H}$] &  $r_{SNR}$ [pc]  & $r_{CS}$ [pc] \\
\tableline
2.0 & 4.13 & 4.59 \\
2.5 & 3.05 & 3.56 \\
3.0 & 2.23 & 2.72 \\
\tableline
\tableline
\end{tabular}
\end{center}
\end{table}
gives radii of our 1-D remnants at $t=1\times 10^4$ yr ($r_{SNR}$), 
and distances between explosion site and CS ($r_{CS}$). As it can be 
seen from Fig.\ 1,
%
%
\begin{figure}
\plotone{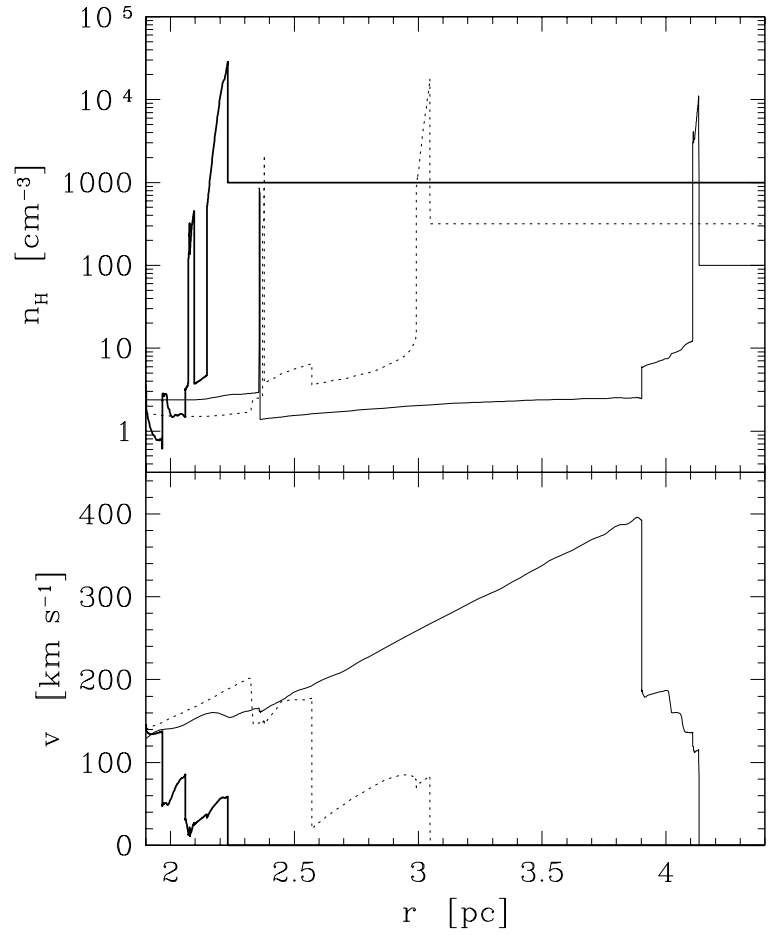}
\caption{Structure of the 1-D supernova remnants at $t=1\times
10^4$ yr. Gas densities and velocities are shown for $\log n_{H}$ =
2 (thin solid), 2.5 (dotted), and 3 (thick solid).}
\end{figure}
models obtained with different ambient densities differ mainly in
their final radii, and width of the shell (the shell is thicker, and
its velocity is lower, in denser ambient media). We will see that the
latter property has a direct implication for accretion onto the CS.

Fig.\ 2
%
%
\begin{figure}
\plotone{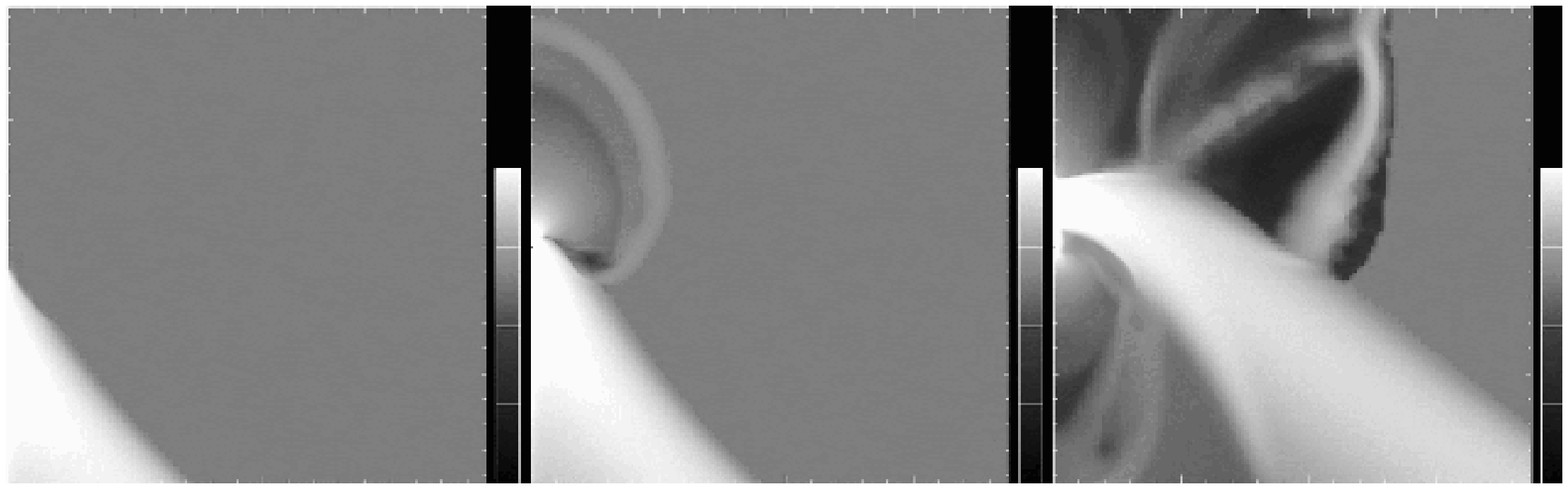}
\caption{Structure of the 2-D model in the vicinity of the Central
Source for $\log n_{H} = 2.5$. The image is 0.2 pc on a side and the
CS is located at the left axis in the middle. The density field in log
scale is shown at $t=12643$ (left panel), 12712 (middle panel), and
$13400$ yr (right panel), with $\log\varrho_{min}=-25$ (black) and
$\log\varrho_{max}=-19$ (white).}
\end{figure}
shows the density distribution in the 2-D model for $\log n_{amb}=2.5$
(a detailed discussion of all cases will be presented in the
forthcoming publication). An early stage of the evolution ($t=12643$
yr), shortly before the material from the shell starts falling into
the CS, is displayed in the left panel in Fig.\ 2. As one can clearly
see, it takes about 2000 years years for CS gravity to completely
deform the SNR shell.  The accretion begins from a tip of the cusp
extending towards the CS, but part of the shell material flows past
the accretor to form an "external bubble" behind it (middle panel,
$t=12712$ yr). We find that this bubble forms unless the width of the
shell is much smaller than the radius of the CS.

During later stages of the interaction the external bubble does not
seem to have a significant influence on the evolution ($t=13400$ yr,
right panel in Fig.\ 2). Instead, we observe a formation of a fast
($v\approx 5000$ km/s) wind blowing from the CS region into the
interior of the remnant.  In the the right panel of Fig.\ 2, this wind
is shocked at a distance of $\approx 1$ pc from the CS to temperatures
$> 10^8$ K with pre-shock densities $n\approx 0.02\,n_{H}$. We note
that the mechanism of wind formation in our model resembles that of a
jet-like flow found by Fryxell, Taam, \& McMillan (1987) in their
study of the axisymmetric flow past a gravitating sphere.

In Fig.\ 3
%
%
\begin{figure}
\plotone{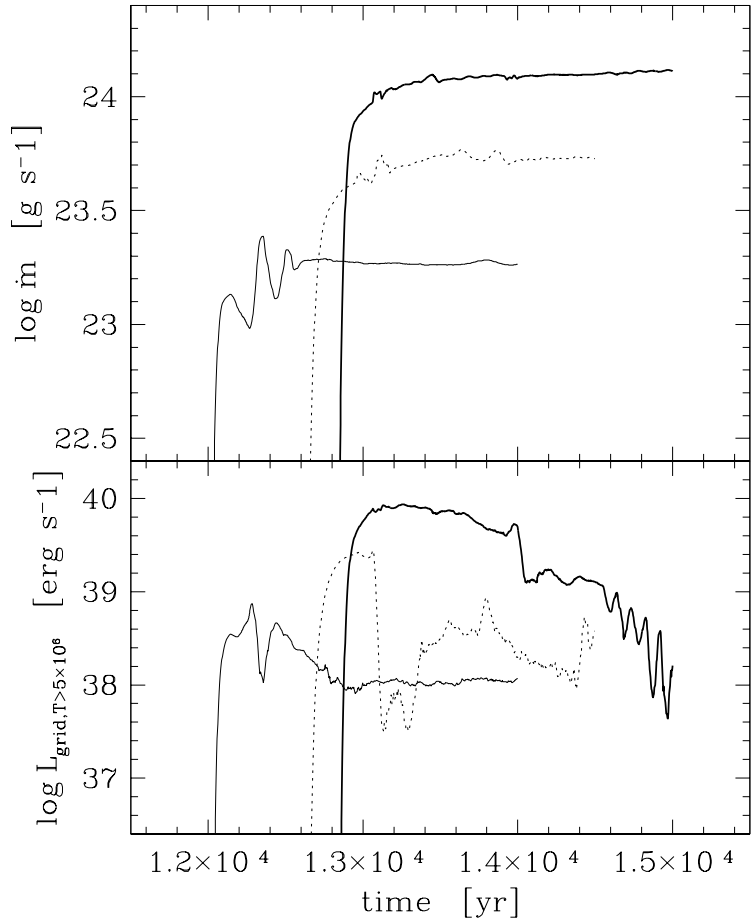}
\caption{Model accretion rates (top panel) and grid luminosities for
$T > 5\times 10^6$ K (bottom panel) during early stages of the
accretion phase. The data is plotted with thin solid, dotted, and
thick solid lines for $\log n_{H}$ = 2, 2.5, and 3, respectively.}
\end{figure}
we show accretion rates and luminosities during the first few hundreds
years of the accretion episode. In all cases the increase in both
accretion rate and luminosity is very rapid. Otherwise, the degree of
correlation between accretion rate and luminosity is rather poor,
presumably due to inadequate resolution in the immediate vicinity of
the CS. As we do not resolve the Schwarzschild radius, our model
provides only lower limits for the luminosity. However, we note that
the model predicts a copious production of energetic emission with the
peak luminosity approaching $10^{39}$ erg/s for the low-density
model. For intermediate and high density the luminosity exceeds the
lower limit ($L_{X} \ga 10^{39}$ erg/s, Murakami et al. 2001) required
to explain the existence of the X-ray reflection nebula in the
Galactic center.

This work was supported in part by the US Department of Energy under
Grant No. B341495 to the Center of Astrophysical Thermonuclear Flash
at the University of Chicago, and in part by the grant 2.P03D.014.19
from the Polish Committee for Scientific Research.

\end{document}